\documentclass[aps, prb,twocolumn,showpacs,superscriptaddress]{revtex4-1}
\usepackage{amssymb,amsmath,bm,graphicx}
\usepackage{color}

\newcommand{\bsigma}{{\boldsymbol \sigma}}

\begin{document}

\title{Intrinsic Anomalous Hall Effect in Type-II Weyl Semimetals}
%
  \author{A. A. Zyuzin}
    \affiliation{Department of Physics,
University of Basel, Klingelbergstrasse 82, CH-4056 Basel, Switzerland}
\affiliation{A.F.Ioffe Physico-Technical Institute, 194021 St. Petersburg, Russia}

\author{Rakesh P. Tiwari}
\affiliation{Department of Physics, University of Basel,
    Klingelbergstrasse 82, CH-4056 Basel, Switzerland}
    
\date{\today}

\begin{abstract}
Recently, a new type of Weyl semimetal called type-II Weyl semimetal has been 
proposed. Unlike the usual (type-I) Weyl semimetal, which has a point-like 
Fermi surface, this new type of Weyl semimetal has a tilted conical spectrum around the Weyl point.
Here we calculate the anomalous Hall conductivity of a Weyl semimetal with a tilted conical spectrum 
for a pair of Weyl points, using the Kubo formula. We find that the Hall 
conductivity is not universal and can change sign as a function of the parameters quantifying the tilts.
Our results suggest that even for the case where the separation between the Weyl points vanishes, tilting of the 
conical spectrum could give rise to a finite anomalous Hall effect, if the tilts of the two cones are not identical.

\end{abstract}

\maketitle

\section{Introduction}
In recent years, condensed matter systems with topologically nontrivial band structures 
have generated a lot of interest. One particularly intriguing topological system is the 
three-dimensional Weyl semimetal\cite{mur07}. In the simplest possible realization
of a Weyl semimetal, there are at least two \textit{distinct} points, called Weyl points, in the first Brillouin zone
, where the conduction and the valence bands touch. These points 
always come in pairs, which are usually protected by some crystalline symmetry, and represent 
a source and a sink of Berry curvature\cite{wan11,bur11a,bur11b,GalBal,zyu12,liu13,yan11,vol09}. 
Experimental realizations of Weyl semimetals have been reported in $\textrm{TaP}, \textrm{NbP}, \textrm{TaAs}, \textrm{NbAs}$\cite{xu15a, lv15a, lv15b, 
she15, yan15, xu15b, zha15}. The topological nature of Weyl semimetals gives rise to 
Fermi arc surface states, the quantum anomalous Hall effect, and chiral-anomaly related negative magnetoresistance~\cite{nie83,zyu12b,hos13,Burkov_rev,son13,zha15b,xio15,hua15}. 

\begin{figure}[t!]
\includegraphics[width=8.0cm,height=8.0cm]{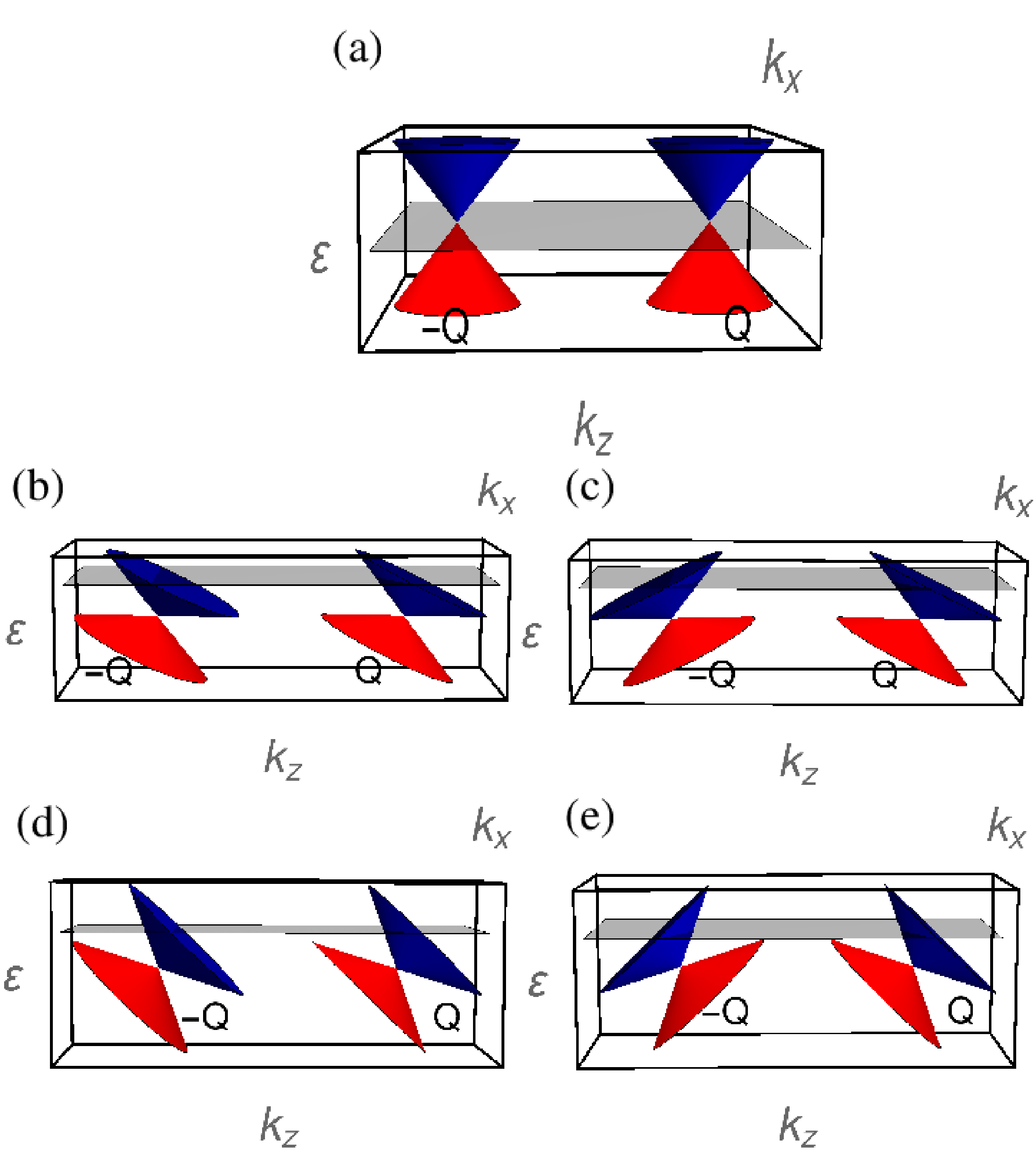}\caption{(color online) The tilted conical spectrum around the Weyl points (WPs). (a) 
Two untilted WPs are located at $\mathbf{k}=(0,0,\pm Q)$ and the low-energy excitations around these points are 
described by Eq.(\ref{eq:Hamkz}), where $C_1=C_2=0$. (b) Tilting the original WPs in the same direction towards $k_z$, see Eq.(\ref{eq:Hamkz}), where $C_1=C_2=-v$. 
(c) Tilting the original WPs in the opposite direction towards $k_z$, see Eq.(\ref{eq:Hamkz}), where $C_1=-C_2=-v$. (d) By increasing the tilts in (b) further, we can reach a situation, where electron and hole like cones of the same WP (located either at $-Q$ or $+Q$) approach the plane $\varepsilon = -\hbar |C_1| (k_z \pm Q)$ (corresponding to $|C_1|\gg v$). 
(e) By increasing the tilts in (c) further, we can reach a situation, 
where electron and hole like cones of the same WP (located either at $-Q$ or $+Q$) approach the plane $\varepsilon = \pm \hbar |C_1| (k_z \pm Q)$ (corresponding to $|C_1|\gg v$).
The grey plane corresponds to the chemical potential $\mu$, measured 
from the WPs, which are located at $\varepsilon=0$. For (a) $\mu=0$, while in (b), (c), (d), and (e) $\mu>0$. We set a cut-off for the dispersion around WPs 
beyond which the excitations are no longer described by a linearized model. 
  \label{fig:tiltconekz}}
\end{figure}

\begin{figure}[t!]
\includegraphics[width=8.0cm,height=8.0cm]{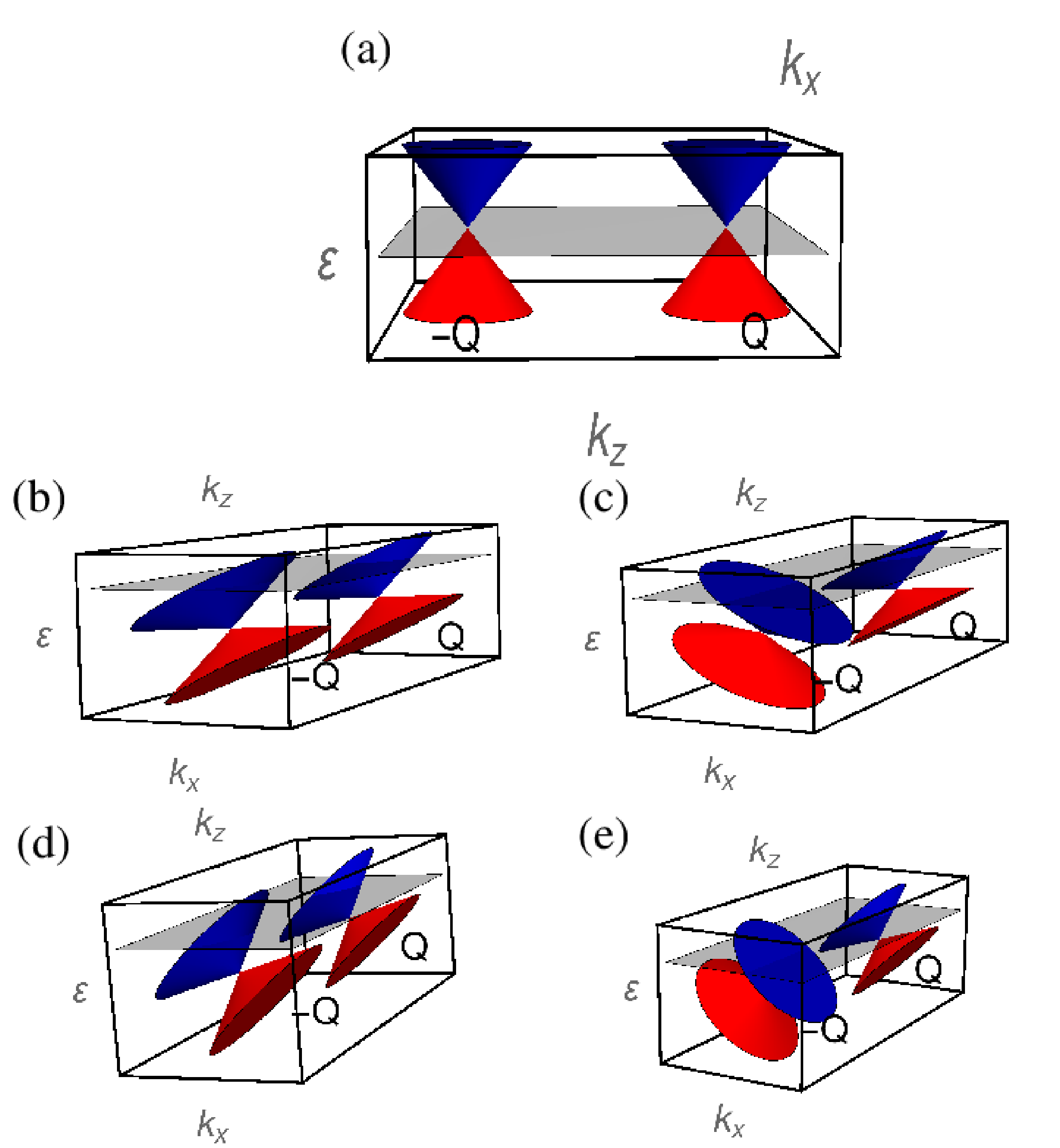}\caption{(color online) 
The tilted conical spectrum around the Weyl points (WPs). (a) 
Two untilted WPs are located at $\mathbf{k}=(0,0,\pm Q)$ and the low-energy excitations around these points are 
described by Eq.(\ref{eq:Hamkx}), where $C_1=C_2=0$. (b) Tilting the original WPs in the same direction towards $k_z$, see Eq.(\ref{eq:Hamkz}), where $C_1=C_2=v$. (c) Tilting the original WPs in the opposite direction towards $k_z$, see Eq.(\ref{eq:Hamkz}), where $C_1=-C_2=-v$.
(d) By increasing the tilts in (b) further, we can reach a situation, where electron and hole like cones of the same WP (located either at $-Q$ or $+Q$) approach the plane $\varepsilon = \hbar |C_1| k_x$ (corresponding to $|C_1|\gg v$).
(e) By increasing the tilts in (c) further, we can reach a situation, 
where electron and hole like cones of the same WP (located either at $-Q$ or $+Q$) approach the plane $\varepsilon = \mp\hbar |C_1| k_x$ (corresponding to $|C_1|\gg v$). The grey plane corresponds to the chemical potential $\mu$, measured 
from the WPs, which are located at $\varepsilon=0$. For (a) $\mu=0$, while in (b), (c), (d), and (e) $\mu>0$. 
We set a cut-off for the dispersion around WPs 
beyond which the excitations are no longer described by a linearized model.
  \label{fig:tiltconekx}}
\end{figure}

Most recently, topological transformation of the Weyl point to a Dirac line or a Fermi surface has been investigated theoretically~\cite{vol14}. Alternatively, it has been shown that a Weyl point can also develop a pair of non-degenerate gapless spheres~\cite{WSM2}. Ideal Weyl semimetal has a conical spectrum and a point-like Fermi surface at the Weyl point. Imagine this conical spectrum \textit{getting} tilted towards some direction in the Brillouin zone (see Figs.~\ref{fig:tiltconekz} and ~\ref{fig:tiltconekx}). 
This tilt can be attributed to strain or chemical doping of the original Weyl semimetal. If this tilt is small enough such that the Fermi surface remains point-like, the 
system is classified as type-I Weyl semimetal. 
However, large tilting of the conical spectrum results in a Lifshitz transition to a new phase classified as type-II Weyl semimetal, 
where the Fermi surface is no longer point-like~\cite{vol14,sol15}, but instead consists of electron and hole pockets, such that the density of states at the Weyl point is \textit{finite}. For a linearized model as shown in Figs.~\ref{fig:tiltconekz} and ~\ref{fig:tiltconekx}, the density of states at the Fermi level depends upon the cutoff 
in momentum space, 
beyond which a linear description of the excitations around the Weyl point is no longer valid. 

It was shown theoretically that WTe$_2$ is a possible candidate for an experimental realization of such a phase~\cite{sol15}. The various transport and thermodynamic properties of type-II Weyl semimetals are very different from those of the type-I Weyl semimetal, partially due to marked differences in the density 
of states of these semimetals at the Fermi level~\cite{sol15}. 
In particular, Ref.~\onlinecite{sol15} points out the anisotropy of the chiral-anomaly related negative magnetoresistance in type-II Weyl semimetal, such that this effect vanishes if the external magnetic field is applied along an axis perpendicular to the direction of tilt of the conical spectrum. 

In this article, we investigate the anomalous Hall effect (AHE) in ferromagnetic type-II Weyl semimetal and 
compare our results with those for type-I Weyl semimetal. AHE describes the contribution to the Hall resistivity of ferromagnetic metals proportional to the magnetization and originates from spin-orbit coupling in this materials. 
There are two contributions to the AHE in ferromagnetic metals, the extrinsic and intrinsic. The so-called intrinsic contribution originates from the 
band structure of the ferromagnetic metal, and the extrinsic contribution from the impurity scattering, see Refs.~\onlinecite{nag10,xia10}. The intrinsic contribution to the AHE was shown to have a direct connection to the presence of nontrivial geometric phase in the electronic 
wave function~\cite{nag10,xia10}. 
Practically, intrinsic mechanism should dominate in the perfect crystals, while extrinsic contribution plays a role in the highly disordered ferromagnets, which makes experimental identification of different sources of AHE 
challenging~\cite{nag10,xia10}.

It was recently demonstrated that the AHE in ferromagnetic type-I Weyl semimetal is purely intrinsic and fully determined by the relative location of the Weyl points. 
This fundamental property of Weyl semimetals is a consequence of the linear dispersion of the excitations close to Weyl points and the existence of non-zero topological charges 
associated with them~\cite{bur14}. Thus, ferromagnetic Weyl semimetals could serve as an ideal platform for the experimental investigation of geometric effects via the anomalous Hall conductivity.

It is thus of theoretical importance to understand if the AHE is also a universal property of ferromagnetic type-II Weyl semimetals.
We find that the AHE in type-II Weyl semimetal depends crucially on the strength and direction of tilts of the conical spectrum. 
We show that in addition to the contribution coming from the relative location of the Weyl points, the tilting gives rise to anisotropic intrinsic Fermi surface contribution.
This makes anomalous Hall conductivity in ferromagnetic Weyl semimetal with tilted Weyl cones indistinguishable from an ordinary ferromagnetic metal, besides the effect of anisotropy. 
We explore two distinct regimes, one where the tilts are in the direction pointing along the line separating the two Weyl points (see Fig.~\ref{fig:tiltconekz}) and the other where the tilts are in a direction perpendicular to this line (see Fig.~\ref{fig:tiltconekx}). For each of these regimes, we consider various possible tilting angles.

The rest of this article is organized as follows. In the next section (Sec. II) we describe our model and the calculation of the AHE in type-II Weyl semimetal. In Sec. III, we present a
discussion of our results, highlighting the key differences between the AHE in both types of semimetals, and finally in Sec. IV, we present a summary of our main results before 
presenting our concluding remarks.

\section{Derivation of principal results}

Consider the simplest kind of Weyl semimetal state with only two tilted Weyl points, separated in momentum space due to violation of the time reversal symmetry in the system~\cite{bur11a}.
We analyze the AHE in two different regimes, the first where the direction of tilt is along the 
direction in which the two Weyl points are separated and the second where the direction of tilt is orthogonal to the direction in which 
the two Weyl points are separated. In both regimes the Hall conductivity is calculated using the Kubo formula.

\subsection{Splitting of Weyl points along the direction of tilt} 
The low-energy model of two Weyl points of opposite chirality, at the same energy, and separated in the momentum space is described by the Hamiltonian
\begin{eqnarray}
H_{1,2}(\mathbf{k}) = \hbar C_{1,2} (k_z\mp Q) \mp \hbar v \bsigma \cdot (\mathbf{k} \mp Q \mathbf{e}_z),
\label{eq:Hamkz}
\end{eqnarray}
where $2Q$ is the distance between the Weyl points in momentum space along $\mathbf{e}_z$, 
$v$ is the Fermi velocity when $C_1=C_2=0$, and $\bsigma$ 
is a vector composed of the three Pauli matrices.
We set $\hbar =1$ throughout the intermediate steps of our calculation and restore it in the final expressions.
The type of the Weyl point is defined by the parameter $C_{\chi}$, $\chi\in\{1,2\}$, such that it is of type-I if $|C_{\chi}|<v$ and type-II if $|C_{\chi}|>v$. In former case the point nodes coexist with 
the electron and hole Fermi pockets~\cite{vol14,sol15}. The one-particle Green functions have the following form
\begin{equation}
G_{1,2}(\omega_n,\mathbf{k}) = \sum_{s=\pm1} \frac{(1-s\bsigma \cdot \mathbf{N}_{\mathbf{k}\mp Q\mathbf{e}_z})/2}{i\omega_n+\mu -C_{1,2} (k_z\mp Q) +s v |\mathbf{k}\mp Q\mathbf{e}_z|},\\
\end{equation}
where $\omega_n$ is the fermionic Matsubara frequency, 
$\mu$ is the chemical potential, and $\mathbf{N}_{\mathbf{k}} = \mathbf{k}/k$ is the unit vector in the direction of wave-vector $\mathbf{k}$.
In what follows we assume for concreteness that $\mu\geqslant 0$.

The anomalous Hall conductivity is given by the zero frequency and zero wave-vector limit of the current-current correlation function 
\begin{eqnarray}\nonumber
\Pi_{ij}(\Omega,\mathbf{q}) &=& T\sum_{\omega_n}\sum_{\chi=1,2}\int \frac{d^3k}{(2\pi)^3} J_{i}^{(\chi)} G_{\chi}(\omega_n+\Omega_m,\mathbf{k}+\mathbf{q})\\
&\times&J_{j}^{(\chi)} G_{\chi}(\omega_n,\mathbf{k})\bigg|_{i\Omega_m\rightarrow \Omega + i\delta},
\end{eqnarray}
where $T$ is the temperature (the Boltzmann constant is set to unity), as:
\begin{eqnarray}
\sigma_{xy} = -\lim_{\Omega\rightarrow 0}\frac{\Pi_{xy}(\Omega,0)}{i\Omega}.
\end{eqnarray}
The current operators are defined as follows
\begin{equation}
J_{i}^{(1,2)} = e  \left(C_{1,2}\delta_{iz} \pm v \sigma_i \right).
\end{equation}
Performing the summation over the fermionic frequencies we obtain $\Pi_{xy}(\Omega,0) = \Pi_{xy}^{(1)}(\Omega,0)+\Pi_{xy}^{(2)}(\Omega,0)$, where 
\begin{eqnarray}\nonumber
&\Pi_{xy}^{(1,2)}(\Omega,0)= \pm e^2 \int_{-\Lambda \mp Q}^{\Lambda \mp Q} \frac{dk_z}{2\pi}\int_0^{\infty} \frac{k_{\perp}dk_{\perp}}{2\pi} \frac{k_z}{k}\frac{2v^2 \Omega_m}{\Omega_m^2+4v^2k^2}\\
&\times
\bigg \{ f(C_{1,2}k_z+vk)-  f(C_{1,2}k_z-vk)
 \bigg\}\bigg|_{i\Omega_m\rightarrow \Omega + i\delta},~~
\end{eqnarray}
$f(E) = (e^{(E-\mu)/T}+1)^{-1}$ is the Fermi distribution function, and $k = \sqrt{k_z^2+k_{\perp}^2}$. 
Here we have introduced a momentum cut-off $\Lambda$ along the $z$-axis. This cut-off is necessary for correct evaluation of the AHE for Weyl semimetals within the linear dispersion model~\cite{gos13}.
For the type-II Weyl semimetal, the momentum cut-off $\Lambda$ is a measure of the density of states due to electron and hole Fermi pockets. 
Taking the zero temperature limit we obtain the anomalous Hall conductivity in the limit of zero frequency $\sigma_{xy} = \sigma_{xy}^{(1)}+\sigma_{xy}^{(2)}$, where 
\begin{eqnarray}\nonumber
\sigma_{xy}^{(1,2)}&=& \mp \frac{e^2}{8\pi^2}\int_{-\Lambda \mp Q}^{\Lambda \mp Q} dk_z
\bigg \{ \mathrm{sign}(k_z) \Theta(v^2 k_z^2 - (C_{1,2}k_z -\mu)^2)\\
 &+& \frac{vk_z}{|C_{1,2}k_z-\mu |}(1-\Theta(v^2 k_z^2 - (C_{1,2}k_z -\mu)^2) )
 \bigg\}.
\label{eq:sigmaxy12kz}
\end{eqnarray}
In the limit when the Fermi energy is at the charge neutrality point, $\mu=0$, we find that the Hall conductivity 
is a non-analytic function of $C_{1,2}$:
\begin{equation}\label{Mu0}
\sigma_{xy} = \frac{e^2 Q}{2 \pi^2 \hbar} \left[ \min\bigg(1, \frac{v}{|C_1|}\bigg)+\min\bigg(1, \frac{v}{|C_2|}\bigg)\right]\frac{1}{2}.
\end{equation}
Here, we have restored the Planck constant $\hbar$. Note that $\sigma_{xy}$ does not depend on $C_{1,2}$ for $|C_{1,2}|<v$, while it becomes non universal at $|C_{1,2}|>v$ and decreases with an increase of $|C_{1}|$ or $|C_{2}|$. 
In both limits $\sigma_{xy}$ is proportional to the distance between the Weyl points\cite{bur11a}. 
The condition $|C_{\chi}| = v$ describes the case when the Weyl cone touches the Fermi level. Thus the non analytic behaviour of the anomalous Hall conductivity 
is related to the van-Hove singularity in the density of states at the Fermi level.

\begin{figure}[h]
\includegraphics[width=\columnwidth]{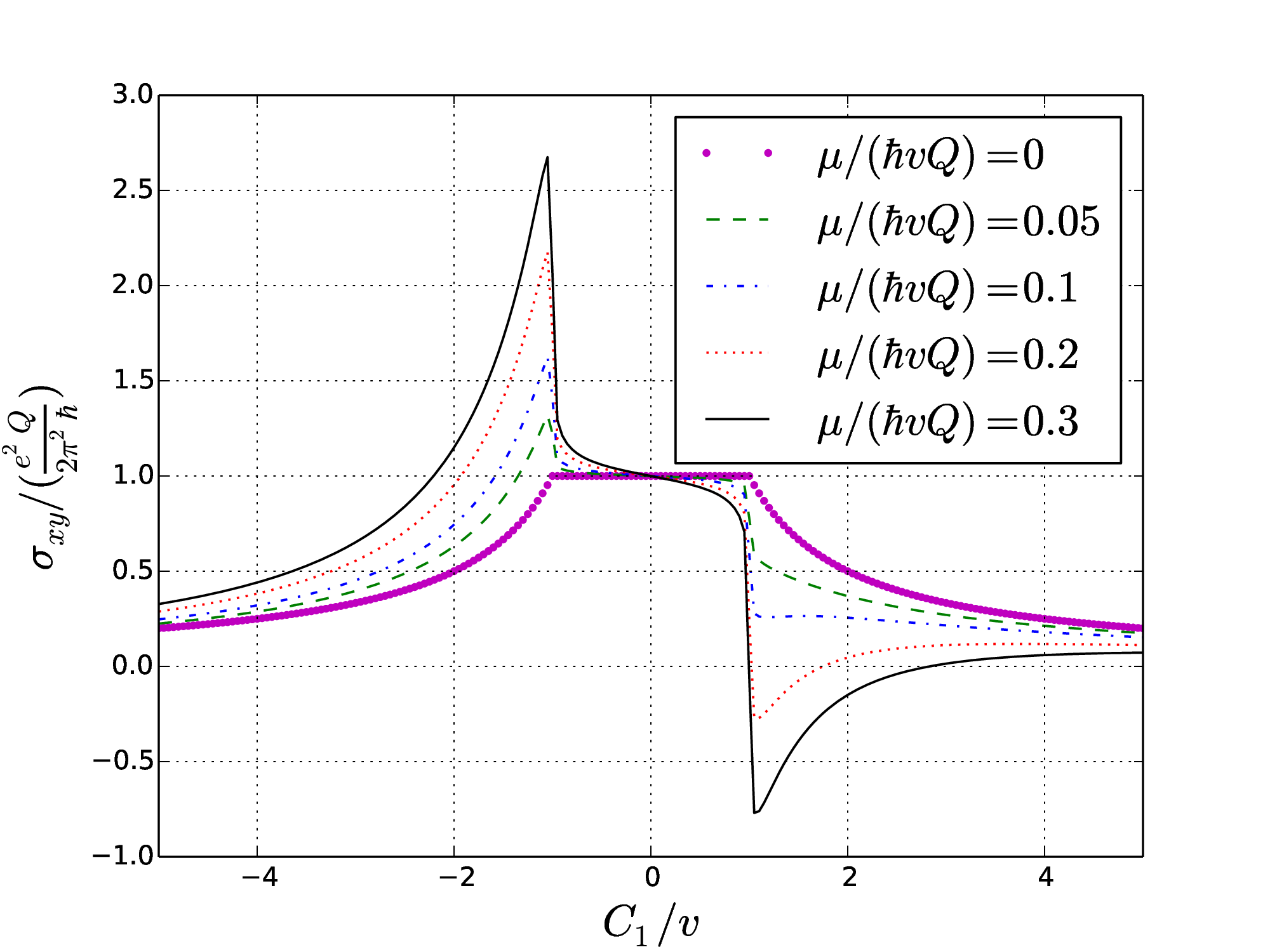}
\caption{\label{fig:Tilt_kz} 
(color online) Normalized anomalous Hall conductivity $\sigma_{xy}/(\frac{e^2 Q}{2 \pi^2 \hbar})$ as a function of the parameter $C_1/v$
for different values of the normalized chemical potential $\mu/(\hbar vQ)$, and fixed $C_2 = -C_1$. Fig.~\ref{fig:tiltconekz}(c) and Fig.~\ref{fig:tiltconekz}(e) show the spectrum for this situation corresponding to $|C_1|=v$ and $|C_1|\gg v$ respectively.}
\end{figure}

The dependence of the Hall conductivity on $C_{1,2}$ at a finite chemical potential is shown in Fig.~\ref{fig:Tilt_kz} . We observe that the
Hall conductivity has a peak or a dip at $|C_{1,2}| = v$ depending on the sign of $C_{1,2}$. The height of the peak or dip diverges logarithmically with the cut-off $\Lambda$~\cite{cutoff} as
\begin{equation}\label{8}
\sigma_{xy} = \frac{e^2}{2\pi^2 \hbar} \bigg[Q + \frac{\mu}{4 \hbar }\bigg(\frac{1}{C}_2-\frac{1}{C_1}\bigg)\bigg(\ln\bigg|\frac{2\hbar v \Lambda}{\mu}\bigg|-1\bigg)\bigg].
\end{equation}
Thus, there is a peak in the Hall conductivity when $C_2 = v$ and $C_1 = -v$ and a dip when  $C_2 = -v$ and $C_1 = v$. 
Interestingly, $\sigma_{xy}$ is finite at $\mu>0$ even if the separation between Weyl points vanishes, $Q=0$. This observation implies that for a finite $\mu$ even a Dirac semimetal ($Q=0$) 
with a tilted conical spectrum such that $C_1\neq C_2$, will show a finite AHE. It is the contribution from the states at the Fermi surface which 
gives rise to a finite value of anomalous Hall conductivity at $\mu>0$.
We find that the anomalous Hall conductivity diverges logarithmically with the cutoff. 
This divergence arises due to the presence of unbounded electron - hole pockets at the Fermi surface, for $|C_{\chi}|\ge v$.

\subsection{Splitting of Weyl points orthogonal to the direction of tilt}
The low-energy model of two Weyl points in the case when the direction of tilt is orthogonal to the point splitting is described by the Hamiltonians
\begin{eqnarray}
H_{1,2}(\mathbf{k}) = \hbar C_{1,2} k_x \pm \hbar v \bsigma \cdot (\mathbf{k} \mp Q \mathbf{e}_z)
\label{eq:Hamkx}
\end{eqnarray}
A calculation similar to the one 
performed in the previous section gives 
\begin{eqnarray}\label{10}
\sigma_{xy}&=& \frac{e^2}{8 \pi^3}\int_{-\Lambda - Q}^{\Lambda - Q}dk_z k_z \int_{-\infty}^{\infty}\frac{dk_x dk_y}{k^3}\\\nonumber
&\times&\bigg[\Theta\bigg(\mu
- C_{1,2}k_x -vk\bigg)-\Theta\bigg(\mu-C_{1,2}k_x +vk\bigg) \bigg].
\label{eq:sigmaxy12kx}
\end{eqnarray}
In the limit $\Lambda \gg \mu$ we recover the expression in Eq.(\ref{Mu0}), which is independent on the Fermi energy. 
Thus, in the case $|C_{\chi}|<v$ the anomalous Hall conductivity does not depend on the parameters $C_{\chi}$ and is given by the
$\sigma_{xy} = \frac{e^2 Q}{2 \pi^2 \hbar}$, while in the case $|C_{\chi}|>v$ the conductivity is not universal and decays with an increase of $|C_{\chi}|$. 
Interestingly, the integral in Eq.~(\ref{10}) vanishes if $Q =0$ even at finite values of the Fermi energy, resulting in a vanishing AHE, which is in contrast to the case when the Weyl cones are tilted along the $z$-axis. 

\section{Discussion and Summary}

Our results indicate that the intrinsic AHE in type-II Weyl semimetal depends crucially on the parameters $C_1$ and $C_2$, and can be used to measure them. 
First we discuss the case, 
where the tilt is in $\mathbf{e}_z$. If both Weyl points are of type-I, ($|C_1| < v$ and $|C_2| < v$) the Hall conductivity for $\mu=0$ is universal (independent of the 
parameters $C_1$ and $C_2$) and is given by $\sigma_{xy}=\frac{e^2Q}{2\pi^2\hbar}$. If one of the points is of type-I and the other of type-II, for e.g., $|C_1| > v$ and $|C_2| < v$, 
the total Hall conductivity for $\mu=0$ is independent of $C_2$ but depends upon $C_1$, and is given by $\sigma_{xy}=\frac{e^2Q}{4\pi^2\hbar}(1+\frac{v}{|C_1|})$. The \textit{decreasing} 
contribution from the type-II Weyl point can be understood as follows. Due to the tilting of the conical spectrum the Hall conductivity now has a contribution 
from both electron and hole like carriers. In fact the contribution of this type-II Weyl point vanishes completely in the limit $|C_1|\rightarrow\infty$ and the total Hall conductivity 
becomes $\sigma_{xy}=\frac{e^2Q}{4\pi^2\hbar}$. 
If both cones are of type-II, the Hall conductivity in general depends upon both parameters $|C_1|>v$ and $|C_2|>v$. A particularly interesting situation is reached if the tilts 
of the two cones are opposite to each other. Fig.~\ref{fig:Tilt_kz} shows the anomalous Hall conductivity in this situation for various values of $\mu$ and $C_2=-C_1$. Note that when we 
have a type-II Weyl semimetal, the Hall conductivity changes sign as a function of $C_1$ (in the right half $C_1 > v$) for large values of $\mu$. The decreasing contribution from the type-II Weyl point
is also observed when the tilt is in $\mathbf{e}_x$ [see Eq.(\ref{eq:sigmaxy12kx})].

It is important to stress that our calculations are performed using a linear dispersion model of the excitations around 
the Weyl points. An unbounded linear dispersion is not realistic for a solid state realization of Weyl semimetal. In a realistic case the linear dispersion will have a cut-off ($\Lambda$) 
in momentum space, beyond which the excitations in Weyl semimetal are no longer described by a linearized model. This cut-off 
is also a measure of the density of states at the Weyl point for a type-II Weyl semimetal. The contribution of these large momentum ($>\Lambda$) states is ignored in our calculation. 
Another crucial point to note is that the conduction and the valence 
bands for the two distinct Weyl points are connected 
by these large momentum states. Figs.~\ref{fig:tiltconekz} and ~\ref{fig:tiltconekx} suggest that the electronic band-structure
of the type-II Weyl semimetal can be drastically modified for large values of the parameters $C_1$ and $C_2$. Thus, in general the contribution of the large momentum states to the AHE will depend 
upon the details of the electronic band-structure of a specific realization of type-II Weyl semimetal and will not be universal. 

Finally, we estimate the value of the anomalous Hall conductivity 
assuming that the typical splitting between the two Weyl points is $Q=(0.01) \textrm{\AA}^{-1}$
, we obtain $\sigma_{xy} = \frac{e^2 Q}{2\pi^2 \hbar} \approx 10 $ per Ohm per cm, which is of the same order as the anomalous Hall conductivity in magnetic conductors\cite{AnomHall}. The dependence of $\sigma_{xy}/(\frac{e^2 Q}{2\pi^2 \hbar})$ on the
Fermi energy is shown in Fig.~\ref{fig:Tilt_kz}. 

So far there has been no experimental evidence for the existence of type-II Weyl semimetal. Thus, a measurement of 
the anomalous Hall effect can help in categorizing the two different types of Weyl semimetals. If the Weyl semimetal is of type-II, the measurement of anomalous Hall effect can provide information about the tilt parameters associated with these type-II Weyl semimetals. 
By applying strain and thereby changing the lattice constant for materials such as HgTe~\cite{rua15}, one can potentially tune 
the tilt parameters. However, our calculation is only applicable to Weyl semimetals, where the splitting of the Weyl points is due to time reversal 
symmetry breaking.

To summarise, we have investigated intrinsic contributions to the AHE in type-II Weyl semimetal within a linear dispersion model using the Kubo formula. 
We found that the tilting of the conical dispersion gives rise to a finite contribution to the intrinsic AHE originating from the Fermi surface. Our findings suggest 
that unlike in type-I Weyl semimetal, where both extrinsic and Fermi surface intrinsic contributions are absent,
such that the AHE is purely intrinsic and fully determined by the separation of the Weyl points
~\cite{bur14}, the 
AHE in type-II Weyl semimetals is not universal and in general 
depends on the parameters quantifying the tilt of the conical spectrum around the Weyl points. The expected extrinsic contribution to the AHE in type-II Weyl semimetal is an 
interesting direction for future research.

\begin{acknowledgments}
  We would like to acknowledge fruitful discussions with A. Burkov, C. Bruder, and V. Zyuzin. The work was 
  financially supported by the Swiss SNF and the
  NCCR Quantum Science and Technology.  
\end{acknowledgments}

%
\end{document}